\documentclass{article}

\usepackage{microtype}
\usepackage{graphicx}
\usepackage{subfigure}
\usepackage{booktabs}
\usepackage{amsmath} 
\usepackage{mathtools}
\usepackage{amssymb}

\usepackage{hyperref}

\usepackage[accepted]{icml2019}

\begin{document}

\twocolumn[

\title{A case study of Generative AI in MSX Sales Copilot: Improving seller productivity with a real-time question-answering system for content recommendation}
\date{\vspace{-0.2in}}
\maketitle

\icmlsetsymbol{equal}{*}

\begin{icmlauthorlist}
\icmlauthor{Manpreet Singh}{MS}
\icmlauthor{Ravdeep Pasricha}{MS}
\icmlauthor{Nitish Singh}{MS}
\icmlauthor{Ravi Prasad Kondapalli}{MS}
\icmlauthor{Manoj R}{MS}
\icmlauthor{Kiran R}{MS}
\icmlauthor{Laurent Bou\'e}{MS}
\end{icmlauthorlist}

\icmlaffiliation{MS}{Microsoft, CX Data Cloud + AI}

\vspace{0.4in}

\begin{abstract}

\vspace{0.4in}

In this paper, we design a real-time question-answering system specifically targeted for helping sellers get relevant material/documentation they can share live with their customers or refer to during a call.  Taking the Seismic content repository as a relatively large scale example of a diverse dataset of sales material, we demonstrate how LLM embeddings of sellers' queries can be matched with the relevant content.  We achieve this by engineering prompts in an elaborate fashion that makes use of the rich set of meta-features available for documents and sellers.  Using a bi-encoder with cross-encoder re-ranker architecture, we show how the solution returns the most relevant content recommendations in just a few seconds even for large datasets.  Our recommender system is deployed as an AML endpoint for real-time inferencing and has been integrated into a Copilot interface that is now deployed in the production version of the Dynamics CRM, known as MSX, used daily by Microsoft sellers.

\vspace{0.4cm}

\end{abstract}
\vspace{0.4in}
]

\icmlcorrespondingauthor{\\ Laurent Bou\'e}{laboue@microsoft.com}
\printAffiliationsAndNotice{}

\section{Introduction}

Generative AI has caught the imagination of the world since ChatGPT was released by Open AI towards the end of 2022.  Leading this trend, Microsoft has announced various Copilot initiatives with the first one being the GitHub Copilot in June 2022 to assist users of Visual Studio in code auto-completion.  Subsequently, Microsoft 365 Copilot was announced in March 2023 to convert user input into content in M365 apps, such as Word, Excel, PowerPoint, Outlook, Teams. Around the same time, Microsoft announced the Dynamics 365 Copilot as the world’s first Copilot in both Customer Relationship Management (CRM) and Enterprise Resource Planning (ERP) applications bringing next-generation AI to every line of business.  The Microsoft Sales Experience team has internally customized the Dynamics~365 Copilot for its CRM system, popularly known as MSX and released MSX Copilot in July 2023 to select sellers. 

The goal of the MSX Copilot is to improve seller productivity and it went live with few skills in July 2023.   A skill is a subject area or knowledge or task awareness for the MSX Sales Copilot. The objective of these skills is to assist the sellers prepare for a meeting with a customer more efficiently.  For example, a seller should be able to ask questions to the Copilot chatbot in natural language about any of the topics covered by the skills and get relevant answers to their questions.  

We have built and deployed the first machine learning based skill for MSX Copilot called ``MSX Content Recommender'' which is the topic of this paper.  Given a seller's query, the purpose of this skill is to return relevant technical documentation drawn from the Seismic content repository~\cite{seismic}.

\section{Related work}
\label{sec:relatedWork}

With the democratization of transformer models~\cite{vaswani2017attention} like BERT~\cite{devlin2018bert}, RoBERTa \cite{liu2019roberta}, Sentence-Bert \cite{reimers2019sentence} and, more recently, Large Language Models (LLMs) like GPT \cite{openai2023gpt4}, most issues involving Natural Language Processing/Understanding have been approached via the angle of text embeddings. 

In this work we focus on semantic search which consists of finding relevant documents from an input query which is not based on keyword matching but rather on contextual understanding.  Because of these recent advancements in~LLMs many search and chatbot applications are moving from lexical-based search to semantic search.  

Unlike classical semantic search problems where one transforms both the user query and the content of the documents into embeddings, we are instead considering the situation where one only has access to meta-feature data about the documents but not their actual content.

Our strategy consists in transforming the metadata about the documents into an English-formed sentence as explained in Section~\ref{sec:prompt}.  This way, we lift ourselves back to the level of asymmetric semantic search.  On one side, we have short user queries and on the other side we have longer textual description of the documents based on their metadata.  In this context, we employ a 2-stage LLM-based architecture to solve the search problem as explained in Section~\ref{sec:architecture}.  Next, we provide performance evaluation both in terms of latency as well as (human) domain experts relevancy estimation in Section~\ref{sec:perf}.  Finally, we conclude with a description of how our system has been integrated as a Copilot skill currently live in the Microsoft Sales Experience (MSX) platform in Section~\ref{sec:copilot}.

\section{Metadata prompt engineering}
\label{sec:prompt}

As mentioned in the previous section, we are considering the situation where we do not have access to the content of the documents but only to their metadata.

In a manner similar to tabular datasets, documents are characterized by a set of features~$\mathcal{F} = \{ f_1 , \cdots , f_n \}$ and any specific document~$c_k$ takes on specific feature values~$\mathcal{F}(c_k) = \{ f_1 (c_k) , \cdots , f_n (c_k) \}$.  

As an example, one may think of a categorical feature~$f_i = \texttt{format}$ that characterizes the format of a document.  Its possible values for a specific document~$c_k$ may be~$f_i(c_k) \in \{ \texttt{PDF}, \texttt{URL}, \dots \} $ depending on the document.  One may also have numerical features~$f_i$ such that~$f_i(c_k) \in \mathcal{R}$.  For example,~$f_i \equiv \texttt{number of views}$ so that~$f_i(c_k)$ counts how many times document~$c_k$ has been viewed in the past.  In this case, we introduce a user-defined function~$g$ that maps numerical features into different categories depending on the numerical value~$f_i(c_k)$ and on the statistics of~$f_i$.  This way, applying~$g$ to $f_i(c_k)$ transforms the numerical features into finite sets of categories.  A specific implementation of~$g$ will be discussed in Section~\ref{sec:ablation}.

If we define the function~$g$ to be the identity for categorical features, we can absorb both categorical and numerical features under the same notation so that~$g \circ f_i(c_k)$ always corresponds to a string value associated with feature~$f_i$ regardless of whether~$f_i$ is a categorical or numerical feature.

Given a document~$c_k$, we concatenate each one its features~$f_i$ with and their string-valued representations~$g \circ f_i(c_k)$ so that~$c_k$ is now translated into a ``prompt'' defined as
\begin{equation}
\text{Prompt}(c_k) \coloneqq \bigoplus_{i=1}^n \Big( f_i \oplus \Big[ g \circ f_i(c_k) \Big] \Big)
\label{eq:prompt}
\end{equation}
where~$\oplus$ stands for string concatenation.  In practice we have considered~$n \approx 20$ features.  An ablation study of including or not including numerical features with a specific implementation of~$g$ is presented in~Section~\ref{sec:perf}

Since all documents and their metadata are available offline, we can apply our prompt engineering function defined in~Eq.~(\ref{eq:prompt}) to all documents and obtain a set of prompts
\begin{equation}
\mathcal{P} = \{ \text{Prompt}(c_1) , \cdots , \text{Prompt}(c_n) \}
\end{equation}
In other words, we have created a new representation of all the documents in the corpus~$\mathcal{C}$ as a set of textual features~$\mathcal{P}$.

\section{Architecture of the LLM-based recommender model}
\label{sec:architecture}

\begin{figure*}[ht]
\begin{center}
\centerline{\includegraphics[width=1.7\columnwidth]{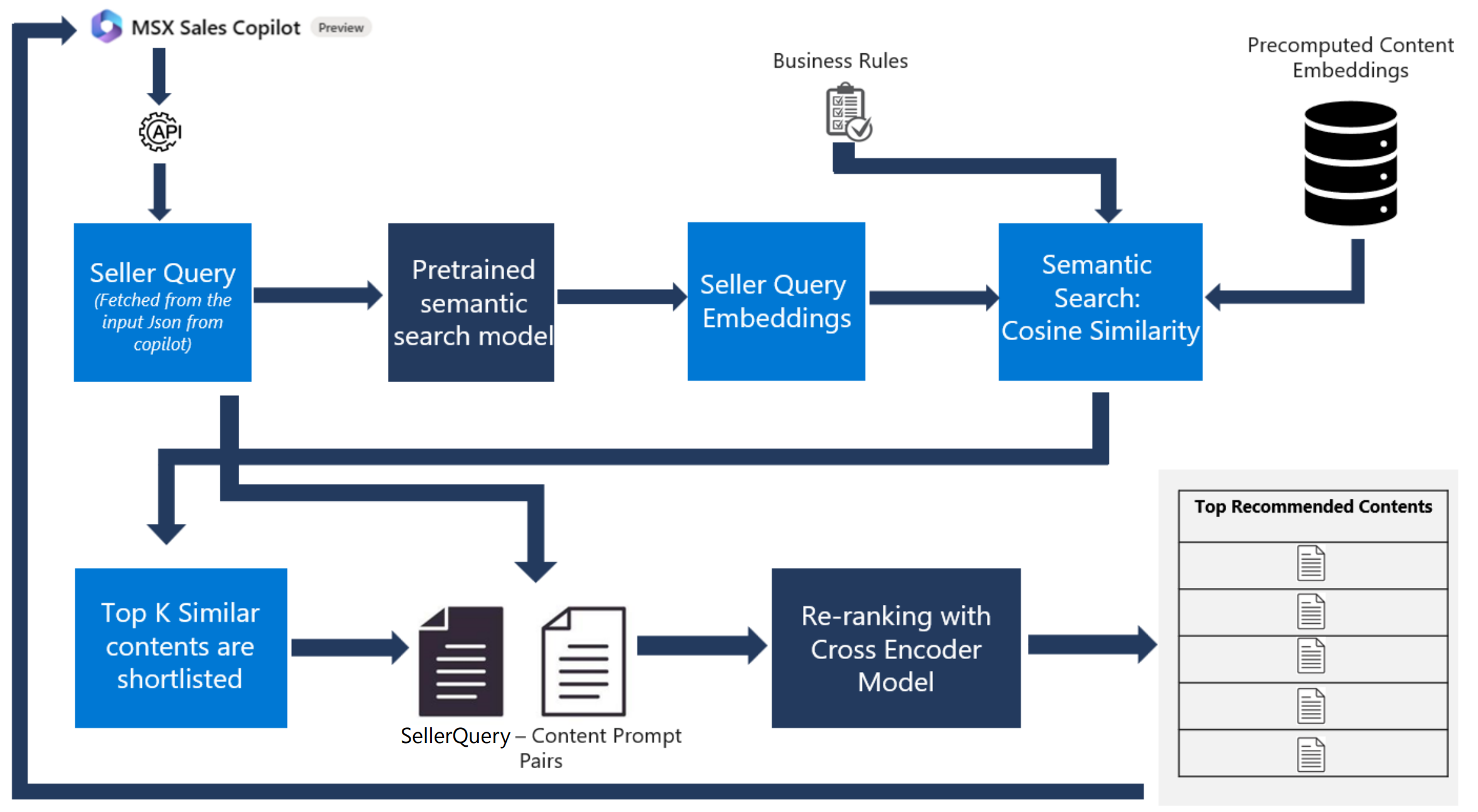}}
\caption{Illustration of the overall 2-stage architecture described in~Section~\ref{sec:architecture}.}
\label{fig:inference}
\end{center}
\end{figure*}

Given a corpus of content $\mathcal{C}=\{ c_1, \cdots ,c_n \}$ (where each~$c_i$ is a document) and a query~$q$ entered into the Copilot interface, the goal of the model is to respond with a list of relevant documents drawn from~$\mathcal{C}$.

Classically, one of the most popular metric to evaluate the relevance of recommender models is the so-called Normalized Discounted Cumulative Gain (NDCG).  Unfortunately this requires to have labeled data for the ground truth relevance score of the search results.  Those are not available in our case and we discuss in the conclusion our future plans in this regard.  Instead we propose to follow a completely unsupervised approach that relies on pre-trained publicly available LLM models.  We define a 2-stage system
\begin{enumerate}
\item the first stage is responsible for quickly retrieving a short list of documents that are potentially good matches to the user query.  This is implemented via a bi-encoder model (see Section~\ref{ref:reterival}).
\item the second stage is responsible for re-ranking the short list produced in the first stage to have more relevant documents appearing first.  This is implemented via a cross-encoder model (see~Section~\ref{ref:crossencoder}).
\end{enumerate}

As mentioned in the previous section, the problem we tackle in this work is that of asymmetric semantic search where a short query is used to match against longer text.  This context of asymmetric question-answering was previously shown to be well-suited for pre-trained publicly available sentence transformers~\cite{reimers-2020-Curse_Dense_Retrieval, reimers2019sentence, modelURL} which are trained on \texttt{MS MARCO} dataset~\cite{bajaj2016ms}.

\subsection{Bi-encoder document retrieval}
\label{ref:reterival}

\begin{figure}[ht]
\begin{center}
\centerline{\includegraphics[width=\columnwidth]{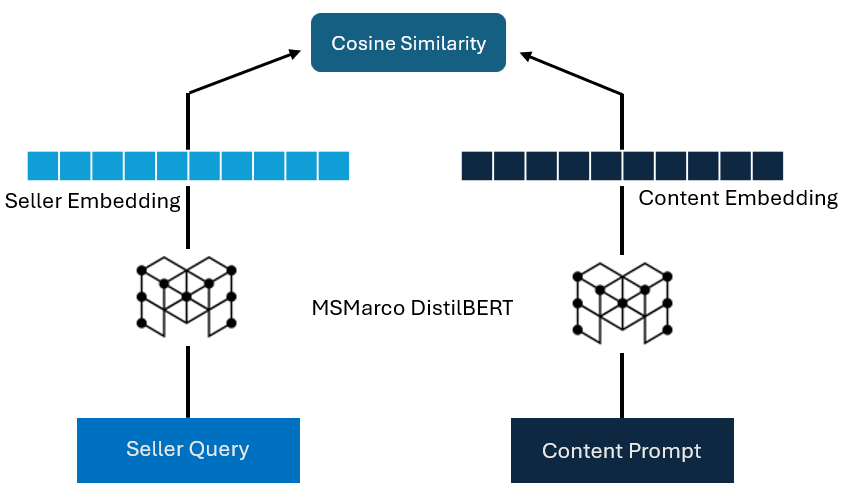}}
\caption{Illustration of the bi-encoder retrieval phase where the embeddings of the metadata prompts~$\mathcal{P}$, pre-computed offline, are compared to the input query~$q$ via cosine similarity.  Note how the two branches operate independently of each other and only connect after the~\texttt{MSMarco DistillBERT} embeddings.}
\label{fig:biEncoder}
\end{center}
\end{figure}

We apply the \texttt{DistillBERT} \cite{sanh2019distilbert} (pre-trained on \texttt{MSMarco} dataset) to all prompts in~$\mathcal{P}$ and cache the resulting set of embeddings~\cite{modelURL}.  Note that this operation may be done completely offline.  At this stage, the prompt embeddings are computed independently from the query embeddings.  This allows for quick retrieval of the top candidates but neglects any kind of cross-interaction between the document prompts~$\mathcal{P}$ and the search query~$q$.
At time of inference (i.e. when a user is invoking a search query), we pass the query~$q$ into the same model to get the query's embedding representation.  Then we use cosine similarity between the pre-computed prompt embeddings and this input query embedding to get the top-$K$ documents
\begin{equation*}
\{ c_{\text{cand}_1} , \cdots , c_{\text{cand}_K}  \} \subset \mathcal{C}
\end{equation*}
that are most closely related to the query as shown in Fig.~\ref{fig:biEncoder}.  Those candidates represent only a small subset of the original corpus~$\mathcal{C}$ of possible documents.

\subsection{Cross-encoder re-ranking}
\label{ref:crossencoder}

\begin{figure}[ht]
\begin{center}
\centerline{\includegraphics[width=\columnwidth]{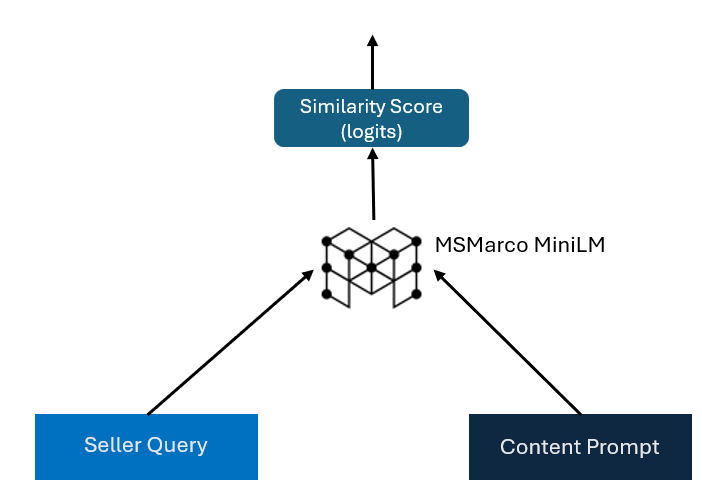}}
\caption{Illustration of re-ranking phase using CrossEncoder where pair of input query text and prompt text are passed to ~\texttt{MSMarco MiniLM} model which produces a score based on similarity between the pair.}
\label{fig:crossEncoder}
\end{center}
\end{figure}

Once the top-$K$ documents have been shortlisted by the bi-encoder document retrieval phase, we now create pairs of the  input query with the metadata prompts corresponding to the candidate documents
\begin{equation*}
\Big[ \big( q, \text{Prompt}(c_{\text{cand}_1}) \big) , \cdots , \big( q, \text{Prompt}(c_{\text{cand}_K}) \big) \Big]
\end{equation*}
Those pairs are then passed together as a unit to the cross-encoder~\texttt{MSMarco MiniLM} model which returns a similarity score for each one of the~$K$ pairs as shown in~Fig.~\ref{fig:crossEncoder}.  Finally, the top-5 ranked documents are returned to the user.

Internally the cross-encoder model uses an attention mechanism between input and prompt to achieve better performance than the retrieval step.  However this performance comes at a cost of higher run times \cite{reimers2019sentence}.  Furthermore, the model cannot be run offline (as we had done previously for the bi-encoder document retrieval phase) since the query~$q$ cannot be known in advance.

To strike a balance between getting accurate documents and low latency, we experimented with different number of top-$K$ candidates shortlisted during the retrieval phase.  We observed that having many candidates~$K \gtrsim 200$ had higher latency and that, on the other end, having few candidates $K \lesssim 20$ would provide too small a dataset to provide relevant documents.  We found that $K=100$ was the best option for re-ranking in a reasonable time while providing relevant documents. 

Latency-wise, the main hyper-parameter of the cross-encoder model we have tuned for performance is the batch size~$b$ which encodes how many samples to load at {\bf inference time}~\footnote{Note that this batch size is at inference time which is slightly unusual in the context of traditional ML models but makes sense in the context of pre-trained LLM inference.  It simply refers to the number of candidates from the top-$K$ returned by the retriever can be loaded simultaneously into the cross encoder.}. We explore the impact of this parameter in detail in Section~\ref{subsection:latency}.

In summary, the overall 2-stage architecture is illustrated in~Fig.\ref{fig:inference}.  Note that such architectures have recently gained more interest due to their ability to combine fast retrieval with powerful and accurate re-rankers~\cite{retrieveAndReRank, zhang2023semantic, reimers2019sentence}.

\section{Performance evaluation}
\label{sec:perf}

To test the performance of the recommender model, we created $31$ queries in collaboration with domain experts from the Seismic catalog~\cite{seismic}.  These evaluation queries are used both for latency and quality evaluation.

\subsection{Inference latency}
\label{subsection:latency}

\begin{figure}[ht]
    \centering
    \includegraphics[width=\columnwidth, height= 0.35\textwidth]{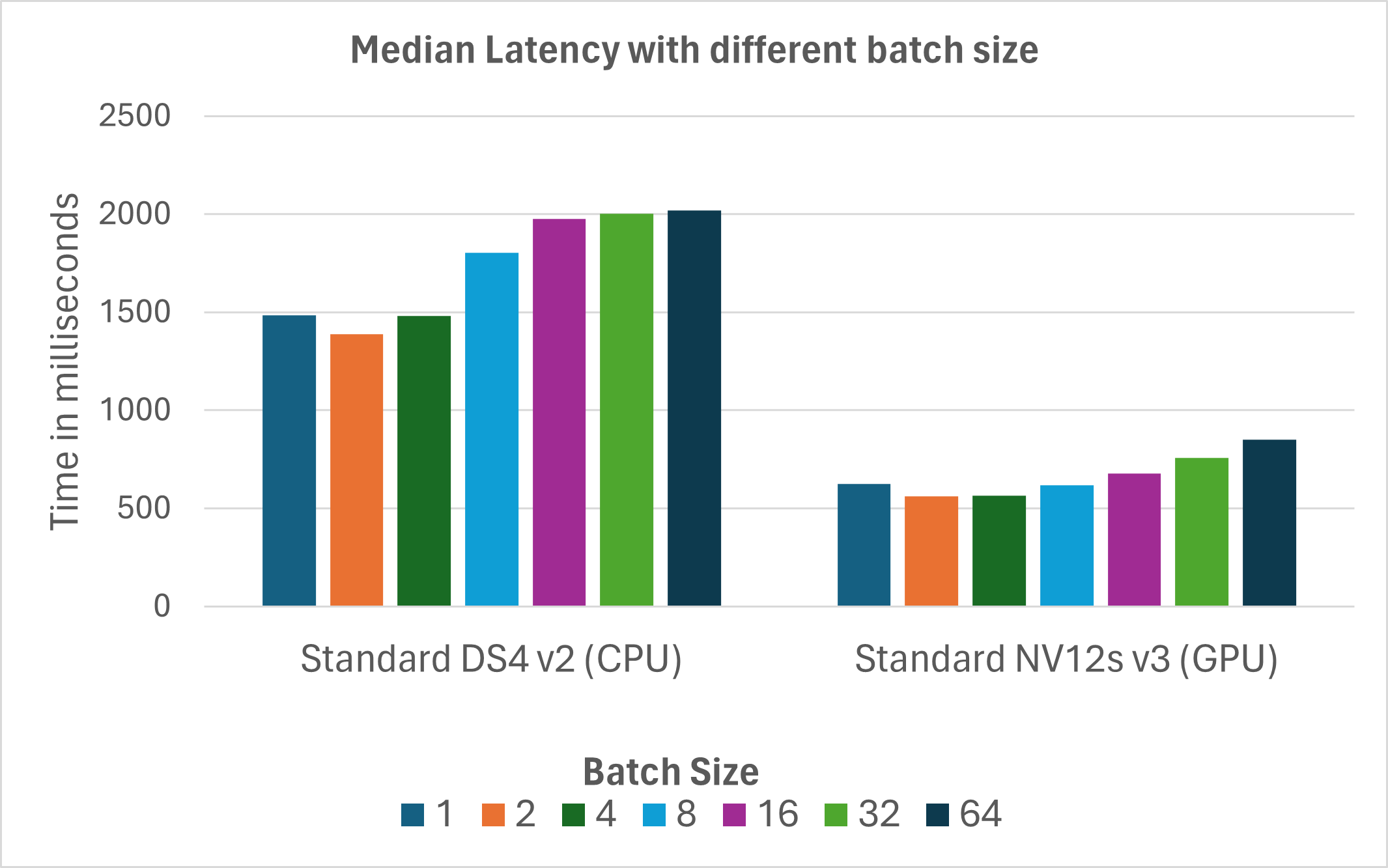}
    \caption{Median latency in milliseconds of different machines on the evaluation queries w.r.t. to batch size~$b$. The latency times represent the overall latency which includes the time taken by bi-encoder to encode the user query to get the embeddings, computing cosine similarity against pre-computed document embeddings, shortlisting the top-100 candidate documents, passing them on to the cross-encoder to re-rank and finally return top-5 documents as described in~Section~\ref{sec:architecture} and~Fig.\ref{fig:inference} (Results for Standard F4s V2 are omitted for the sake of clarity but the actual numbers are reported in Table~\ref{table:inference_latency}.)}
    \label{fig:latency_median}
\end{figure}

\begin{table*}[h]
    \small
    \centering
   \begin{tabular}{|c|c|c|c|c|c|c|c|}
     \hline
           & $b=1$ & $b=2$ & $b=4$ & $b=8$ & $b=16$ & $b=32$ & $b=64$  \\
     \hline
     \textbf{Standard F4s v2} & $4.17 \pm 1.41$ & $4.89 \pm 2.08$ & $5.85 \pm 2.72$ & $7.14 \pm 3.36$ &  $8.03 \pm 3.32$& $8.84 \pm 3.73$ & $10.22 \pm 4.16$ \\
     \hline
     \textbf{Standard DS4 v2} & $1.53 \pm 0.41$ & $1.49 \pm 0.54$ & $1.58 \pm 0.68$ & $1.81 \pm 0.75$ &  $1.96 \pm 0.71$& $2.11 \pm 0.73$ & $2.34 \pm 0.81$ \\ 
     \hline
     \textbf{Standard NV12s v3} & $0.64 \pm 0.07$ & $0.54 \pm 0.13$ & $0.55 \pm 0.18$ & $0.60 \pm 0.19$ &  $0.66 \pm 0.20$ & $0.74 \pm 0.21$ & $0.83\pm 0.25$ \\
     \hline
\end{tabular} 
\caption{Average inference latency with standard deviation as captured on different machines and for different values of the batch size~$b$. (Times are in seconds)}
\label{table:inference_latency}
\end{table*}

As we need to respond to user queries in real-time, we present in this section a study of the inference latency of the model described in~Section~\ref{sec:architecture}.  Mostly we focus on the impact of the batch size~$b$ in the context of the time-consuming cross-encoder phase of the model.  For the purpose of this experiment we used three different types of Virtual Machines on Azure:
\begin{itemize}
\item Standard F4s v2 - 4 Cores, 8 GB RAM, 32 GB disk
\item Standard DS4 v2 - 8 Cores, 28 GB RAM, 56 GB disk
\item Standard NV12s v3 - 12 Cores, 112 GB RAM, 336 GB disk, 1 Nvidia Tesla M60 8 GB vRAM
\end{itemize}
We show in~Fig.~\ref{fig:latency_median} the median latency over the evaluation queries for different values of~$b$. In addition, Table~\ref{table:inference_latency} shows average latency with standard deviation over the evaluation queries with different values of the batch size~$b$. We notice that:
\begin{itemize}
\item For CPU machines, as we use the machine with more available cores the overall latency is reduced.  However, as expected using a machine with single GPU performs significantly faster than the CPU machines we tested.
\item As the batch size increases, the overall inference time also increases as shown in Fig.~\ref{fig:latency_median} (which shows the median value) and Table~\ref{table:inference_latency}.
\end{itemize}

As mentioned previously, our model is expected to return responses to a query in real time. So saving a few hundred milliseconds in latency is crucial. This experiment shows that~$b$ can have a significant impact on the latency.  In conclusion, it is best to choose~$b = 2$ or~$b=4$ for the VMs available to us.

\subsection{Human annotators for quantitative relevancy evaluation}

We passed the evaluation queries to our recommender model and presented the top-5 results for each query to four human annotators who were asked to rate the quality of the responses on a scale from~0 to~5.  

Given a query~$q$, a score of~0 indicates that none of the responses are relevant whereas a score of~5 indicates that all~5 responses are relevant to the query.  Clearly any score in between quantifies the proportion of responses that are deemed relevant by the human annotators.  This process of human verification and assessment of the quality of the model yields~4 independent scores ranging from~$0$ to~$5$ for all queries.  We present a summarized analysis in~Table \ref{table:self_evaluation}. This shows that most queries were judged as ``relevant''.  In fact, the annotators also pointed out that those few queries that fell under the category of ``not relevant'' were either too verbose or too generic so that no correct documents could be picked out from the catalog anyway.

We also considered the average score of each human annotator which ranged from 2.74 to 3.26 to verify that no annotators were overly positive or negative in their evaluation thereby confirming that the results did not suffer from bias.  

\begin{table}[ht]
    \centering
   \begin{tabular}{|c|c|c|}
     \hline
     \textbf{Bucket} & \textbf{Score} & \textbf{Number}  \\
      & \textbf{Range} & \textbf{of Queries} \\
     \hline
     Relevant & $ [3.5-5]$ & $15/31$ \\
     \hline
     Somewhat relevant & $[2-3.5)$ & $9/31$  \\
     \hline
     Not relevant & $[0-2)$ & $7/31$ \\
     \hline
\end{tabular}
\caption{The average scores of the human annotators are bucketed into~3 different classes depending on their numerical values. Most of the queries which fell into the \textit{relevant} bucket were short queries specifying what kind of documents or products they were looking for.   The queries which fell into \textit{Not relevant} bucket were verbose and generic in nature.  The human annotators also noticed that in some cases even if query was overly verbose but if it had the right keywords our model was able to return some relevant contents.}
\label{table:self_evaluation}
\end{table}

\subsection{Ablation study}
\label{sec:ablation}

\paragraph{Architecture}  In this section, we demonstrate by example that the 2-stage architecture, although more complex and costly than a simple 1-stage (i.e. retiever only without cross-encoder), returns noticeably more relevant documents.

For example, consider the query \\ \\ \texttt{What are contents available for Dynamics 365? Give a PDF format document.} \\ \\  The retriever returns only~$2$~PDF documents in its top-5 similarity items.  Although the query specifically asks for~PDF documents, the remaining documents are in~OFT and~PPTX formats.  On the other hand, if we allow the retriever to return more~$K=100$ documents, we are bound to increase the recall rate of~PDF documents although all except~2 would still fall outside of the top-5.  This is where the second stage cross-encoder acts as a re-ranker that tightly couples the query with those~$K=100$ documents so that more attention can be given to the~PDF aspect of the query.  In fact, the cross-encoder successfully re-orders the~$K=100$ candidates so that all top-5 documents are in the correct format. 

In order to confirm and make a more quantitative comparison between the full 2-stage system (bi-encoder document retrieval + cross-encoder re-ranking) and a simple~1-stage system (composed of the bi-encoder only), we have submitted~10 of the evaluation queries for human expert verification on the quality of the top-5 results returned by both architectures.  Unsurprisingly, this has revealed that~$90\%$ of the queries received more (or at least equally) relevant recommendations using the full 2-stage architecture.

\paragraph{Numerical features prompt engineering}

A large fraction of the metadata associated with the documents comes in the form of numerical features.

As explained in~Section~\ref{sec:prompt}, those features can also be integrated into the prompt representation of the documents by introducing a function~$g$ that converts numerical values into different categories.

Given a numerical feature~$f_i$ and its value~$f_i(c_k) \in \mathcal{R}$ for document~$c_k$, we defined the mapping function~$g$ as \[ g \circ f_i(c_k) = \begin{cases}
  \texttt{high} & \mbox{if $\,\,\, f_i(c_k) > p_{85}$} \\
  \texttt{medium} & \mbox{if $\,\,\, p_{65} < f_i(c_k) < p_{85} $} \\
  \texttt{low} & \mbox{if $\,\,\, f_i(c_k) < p_{65}$} \\
  \texttt{zero} & \mbox{if $\,\,\, f_i(c_k) = 0$}
  \end{cases}
  \]
where~$p_r$ stands for the~$r^\text{th}$ percentile of~$f_i$ as estimated over all the documents in~$\mathcal{C}$.

Similarly to the~PDF example in the paragraph above, our human annotators confirmed that incorporating those numerical features did have a significant impact on the relevance of the results.  For example, the results returned in response to a query specifically asking for documents with a ``high'' number of views will often miss that detail in the top-5 results of the retrieval phase but correct itself thanks to the second stage cross-encoder re-ranker.

\section{Integration into Copilot}
\label{sec:copilot}

Our model has been successfully integrated into the MSX Sales Copilot.  It is being refreshed on a weekly basis to accommodate changes/latest updates in meta-properties of the documents.

\subsection{Technical details}

Our recommendation model is one of the few skills that has been integrated into MSX Copilot by using Semantic Kernel \cite{semanticKernel}. Semantic Kernel is an open source SDK used to integrate LLMs (or AI services) into applications. When a query is typed in the Copilot chat window, the planner in Semantic Kernel is invoked. Planners are used to generate plans and invoke the respective skills based on the context of the query. Skills (like our recommendation models) are used to perform certain specific functions. Based on the query and its context, the planner may decide to invoke our recommendation skill. This skill then makes an API call to our model which is deployed on Azure Machine Learning (AML) endpoint. Once the AML endpoint is called, our recommendation model processes that query as described in Section \ref{sec:architecture} and returns the top-5 most relevant documents to the chat window as illustrated in the following section.  This entire process takes a few seconds with the majority of it spent by the planner only.

\subsection{How it looks}

\begin{figure}[h]
\begin{center}
\centerline{\includegraphics[width=0.7\columnwidth]{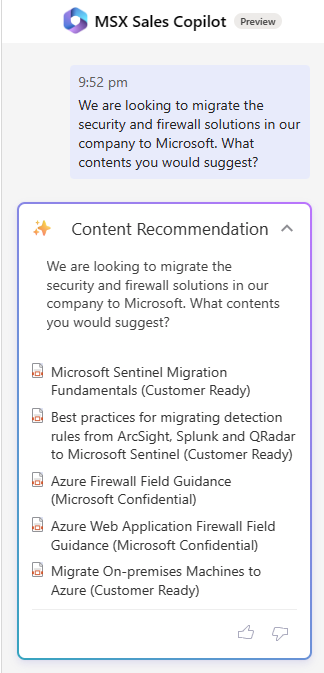}}
\caption{Example of the current UI integration of our model into the production MSX Copilot.}
\label{fig:copilotUI}
\end{center}
\end{figure}

A snapshot of the actual User Interface in MSX is shown in~Fig.~\ref{fig:copilotUI}.  In blue, we see the query from the seller. This along with business rules are sent to our model. The purple outlined box is the response with top-5 documents returned by our model. Each of these has a link to the document which the seller can use to open it and share with the end customer using livesend links. 

\section{Conclusion}

The recommender model presented here is the first to be integrated as a skill into the production MSX Copilot.  The initial feedback from the sellers is encouraging in the first two months since launch.  Compared to their previous experience where sellers had to rely on a sub-optimal filter-based search system hosted externally on the Seismic website (as opposed to directly embedded into the MSX interface), this new real-time question-answering recommender system is seen as a tremendous improvement by the sellers' community as they have reported in satisfaction surveys.  For example, when asked: {\it ``Is the content recommendation skill relevant to the seller's daily tasks?''}, the received score was~4 out of~5.  Similarly, when asked to rate the relevancy of the recommended documents, sellers gave it a rating of 3.7 out of~5.  Moving forward, we are continuously gathering their feedback for further improvement of the model such as integrating more context based on the seller, opportunity etc...  From a technical perspective, the model innovated compared to classical semantic search in the sense that we do not use the content of the documents to recommend but only their metadata.

Our choice of~$\approx 20$ features (see~Section~\ref{sec:prompt}) is limited by the available meta-features available for documents from the Seismic catalog~$\mathcal{C}$ itself.  Generalizing to other corpus of content, it is possible that more features would be available which would presumably help increase the relevance of the recommendations.  Nonetheless, there are other factors that may limit the number of features one may consider.  From an engineering perspective the model needs to operate close to real-time with strict latency requirements.  Adding more features would translate into longer response times and a quality/quantity trade-off on the number of features would need to be found.  Additionally, from a modelling perspective, the MS Marco LLM we used in the current version of the model has a context length of 512 tokens.  With 20 features, this gives each feature about 25 tokens from which a good English description of the feature has to be generated.  Adding more features would reduce the average number of available tokens per feature highlighting the need for another kind of trade-off here as well.

As programmatic access to the actual content of the Seismic documents themselves is not currently available, we limited the current version of the model to be based on meta-features about the documents only.  For other corpus where such access is available, introducing content as a new feature set will raise new and challenging questions.  For example, the documents in the Seismic catalog are cluttered across multiple formats (PowerPoint presentations, PDFs, Word documents…) and one will have to consolidate together multi-modal features (text, images, audio recordings...) into unified embeddings.  We reserve this for a future version of the model once programmatic access to the content of the documents is resolved.

Ultimately, our intentions are to build a content recommendation search and question-answering system that sellers can fully leverage even in customized contexts.  For example, we are exploring how to make the recommendations specific to the opportunities that sellers are invested in.  Sales opportunities are a crucial component in nurturing early customers and more customization in this area is of potential high impact.  Additionally, we intend to incorporate the actual text/images of the documents into the embeddings as well as enrich the query-side prompts with more features specific to the seller asking a question.  Finally, we wish to collect more user feedback as a way of proxied labeled data so we can start incorporating more supervision via metrics such as normalized Discounted Cumulative Gain (nDCG) into the model.  

\section{Acknowledgements}

We thank our colleagues in CX Data for their feedback and support. In particular, we thank the CX Data ML Review board (Vijay Agneeswaran, Daniel Yehdego, Ivan Barrientos) for reviewing the model and giving us useful inputs. We thank the product managers from SPS Team (Binh ``Binnie'' Tran, Manish Prabhakar) who helped us with their domain knowledge on the Seismic platform and the content space. We thank the Copilot engineering team, especially Yash Pal, for providing support in deploying this model to MSX~Copilot platform.

\bibliographystyle{ieeetr}
\bibliography{MainText}

\end{document}